\pdfoutput=1

\documentclass[11pt]{article}
\usepackage{graphicx}
\usepackage{enumitem}

\usepackage{kRAIG}

\usepackage{times}
\usepackage{latexsym}

\usepackage[T1]{fontenc}

\usepackage[utf8]{inputenc}

\usepackage{microtype}

\usepackage{inconsolata}
\usepackage{listings}
\usepackage[most]{tcolorbox}
\usepackage{float}

\lstdefinelanguage{prompting}{
  morekeywords={Prompt, Response, User, Assistant},
  sensitive=true,
  morecomment=[l]{--},
  morestring=[b]",
}
\lstdefinestyle{mylststyle}{
  basicstyle=\ttfamily\small,
  keywordstyle=\color{blue}\bfseries,
  commentstyle=\color{green!50!black}\itshape,
  stringstyle=\color{violet},
  showstringspaces=false,
  tabsize=2,
  breaklines=true,  
  columns=fullflexible, 
  keepspaces=true,
}
\tcbset{
  myboxstyle/.style={
    boxrule = 0pt,
    toprule = 4.5pt, 
    enhanced,
    fuzzy shadow = {0pt}{-2pt}{-0.5pt}{0.5pt}{black!35}, 
    fontupper = \ttfamily\small, 
    boxsep = 5pt, 
    left = 1pt, 
    right = 1pt, 
    top = 1pt, 
    bottom = 1pt, 
  }
}
\title{kRAIG: A Natural Language-Driven Agent for Automated DataOps Pipeline Generation}

\author{Rohan Siva\textsuperscript{*}\thanks{\quad Equal contribution.}\thanks{\quad Work done while at Cisco.} \\
  Cisco \\
  \texttt{rohansiva@utexas.edu\quad} \\\And
  Kai Cheung\textsuperscript{*} \\
  Cisco \\
  \texttt{kacheun2@cisco.com} \\\And
  Lichi Li \\
  Cisco \\
  \texttt{licli@cisco.com} \\\And
  Ganesh Sundaram \\
  Cisco \\
  \texttt{gsundara@cisco.com} \\}

\begin{document}
\maketitle
\begin{abstract}

Modern machine learning systems rely on complex data engineering workflows to extract, transform, and load (ELT) data into production pipelines. However, constructing these pipelines remains time-consuming, and requires substantial expertise in data infrastructure and orchestration frameworks. Recent advances in large language model (LLM) agents offer a potential path toward automating these workflows, but existing approaches struggle with under-specified user intent, unreliable tool generation, and limited guarantees of executable outputs.

We introduce kRAIG, an AI agent that translates natural language specifications into production-ready Kubeflow Pipelines (KFP). To resolve ambiguity in user intent, we propose ReQuesAct (Reason + Question + Act), an interaction framework that explicitly clarifies intent prior to pipeline synthesis. The system orchestrates end-to-end data movement from diverse sources and generates task-specific transformation components through a retrieval-augmented tool synthesis process. To ensure data quality and safety, kRAIG incorporates LLM-based validation stages that verify pipeline integrity prior to execution. 

Our framework achieves a 3x improvement in extraction and loading success and a 25\% increase in transformation accuracy compared to state-of-the-art agentic baselines. These improvements demonstrate that structured agent workflows with explicit intent clarification and validation significantly enhance the reliability and executability of automated data engineering pipelines.
\end{abstract}

\section{Introduction}
\begin{figure*}[t]  
    \centering
    \includegraphics[width=\textwidth]{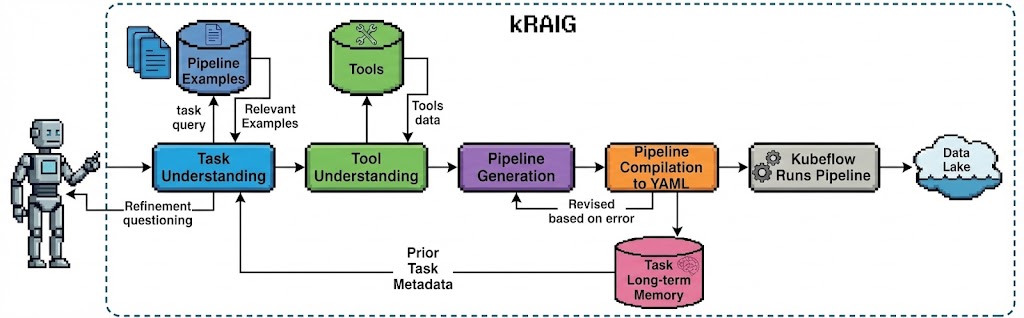}
    \caption{kRAIG Architecture}
    \label{fig:kRAIG-arch}
\end{figure*}

Mapping natural language specifications to executable programs has been an active area of research in semantic parsing and program synthesis \cite{bisk2020experiencegroundslanguage}. While this foundational work established key techniques for language-to-code generation, the landscape of production data systems has evolved significantly. Modern ML pipelines now involve complex orchestration across heterogeneous data sources, streaming platforms, and cloud infrastructure that were less prevalent when these methods were developed.

End-to-end machine learning workflows consist of multiple interdependent stages, including data ingestion, preprocessing, validation, model training, and evaluation. The early stages, data ingestion, transformation, and validation, require substantial manual effort, involving tailored pipelines that must be maintained and updated before model training can begin \cite{chen2025llmsmeetapidocumentation}.

To address these challenges, we introduce k-Retrieval Augmented Integration Governor (kRAIG), an AI agent designed to automate the data processing phase of the machine learning lifecycle. Recent work has shown that large language models can be extended into agentic systems capable of multi-step reasoning and tool invocation \cite{wang2023survey_lm_dialogue,schick2023toolformer}. Building on these capabilities, kRAIG provides a conversational interface where users specify data integration and transformation requirements in natural language, which the system translates into executable data pipelines.

kRAIG is built on top of Kubeflow Pipelines and automatically generates containerized pipeline components to implement user-specified data tasks. Because Kubeflow executes on Kubernetes, kRAIG can leverage distributed compute resources and integrate data processing tool that can be containerized, enabling deployment across large-scale data engineering environments.

In contrast to existing agentic systems that target a wide variety of tasks, such as those evaluated in ELT-Bench, kRAIG focuses on specializing in composable data engineering pipelines \cite{jin2025eltbenchendtoendbenchmarkevaluating}.
 This shift aims to reduce the manual burden of data preparation while preserving the robustness and scalability required in modern machine learning environments.  This initial version of kRAIG focuses on a subset of tools and tasks that are representative of real-world enterprise data engineering workflows.

\subsection{Methodology Comparison}
Existing agent frameworks, such as SWE-Agent and Spider-Agent, are based on the \textbf{ReAct} (Reason + Act) framework that utilizes a recurrent cycle of:
    \begin{enumerate}
        \item	\textbf{Thought} – internal reasoning about the current state
        \item	\textbf{Action} – a concrete step (e.g., call a tool, write code, issue a command)
        \item	\textbf{Observation} – the result of that action
        \item	\textbf{Repeat} until the task is solved
    \end{enumerate}
However, prior studies have shown that purely reason-then-act agent frameworks can exhibit unstable behavior under under-specified or execution-heavy tasks, often failing to recover from early incorrect assumptions \cite{yao2022react}.

We propose a modified interaction framework, termed \textbf{ReQuesAct} (Reason + Question + Act), which explicitly incorporates intent clarification as a step in the agent loop:
    \begin{enumerate}
        \item	\textbf{Thought} – internal reasoning about the current state
        \item   \textit{\textbf{Question}} - intent-clarifying question
        \item	\textbf{Action} – a concrete step (e.g., call a tool, write code, issue a command)
        \item	\textbf{Observation} – the result of that action
        \item	\textbf{Repeat} until the task is solved
    \end{enumerate}

\section{Task understanding}
We adopt a hybrid neurosymbolic approach to task interpretation in which kRAIG engages in a multi-turn dialogue to clarify user intent. As illustrated in Figure~\ref{fig:questioning-schema}, the system incrementally asks clarifying questions to refine the task specification. Rather than requiring users to manually condense their requests into predefined templates, kRAIG analyzes the conversation to extract the structured specification needed to synthesize a data pipeline, reducing user effort compared to traditional template-based systems.

Empirically, we found that without inference-time steering strategies, evaluated models frequently failed to generate appropriate clarifying questions for complex data tasks, consistent with observations reported in QuestBench \cite{li2025questbenchllmsaskright}.
  Once task intent is sufficient, kRAIG queries a retrieval-augmented generation (RAG) component to retrieve the most relevant pipeline examples.  We intentionally restricted the number of example pipelines to evaluate few-shot generalization. With the few pipeline examples kRAIG was able to achieve success in completing our internal tasks.  Rather than increasing the number of example pipelines, our approach emphasizes few-shot generalization from a small set of representative examples.

\subsection{Interface Example}
\begin{figure}[H]
    \centering
    \vspace{-1em}
    \begin{tcolorbox}[myboxstyle]
    \begin{lstlisting}[language=prompting, style=mylststyle]
Assistant: Where should the data from the ELT-Bench GitHub repository be stored?

User: git clone https://github.com/uiuc-kang-lab/ELT-Bench

A: Where should the data from the ELT-Bench GitHub repository be stored?

User: s3 name it cve-bench-new

A: What kind of data transformations would you like to perform on the GitHub repository data before storing it in the S3 bucket?

User: None, actually change the name of the s3 to elt-bench-new
    \end{lstlisting}
    \end{tcolorbox}
    \caption{Example of questioning schema}
    \label{fig:questioning-schema}
\end{figure}

\section{Dynamic tool discovery and on-demand tool synthesis}
\begin{figure}[H]
    \centering
    \includegraphics[width=0.8\linewidth]{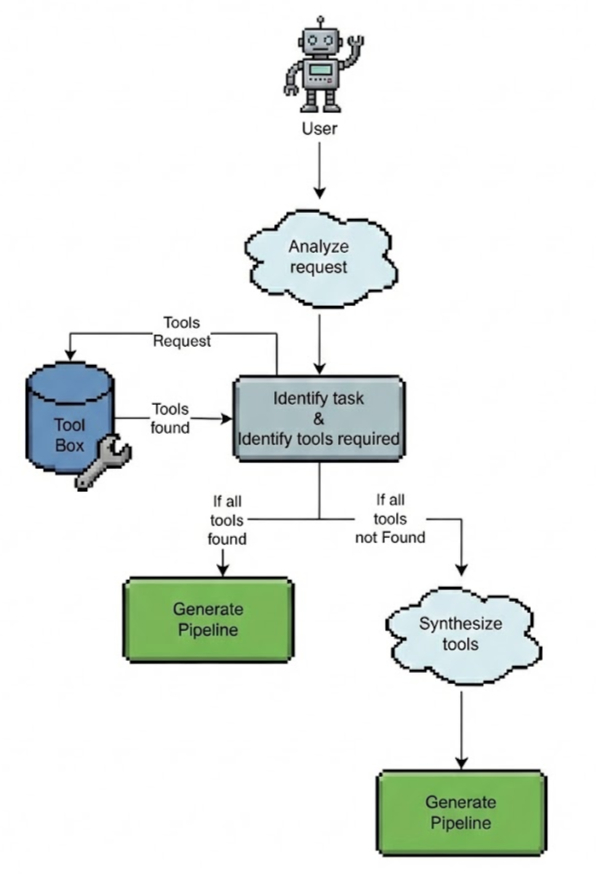}
    \caption{kRAIG tools flow}
    \label{fig:tool_flow}
\end{figure}

After completing the initial task interpretation, kRAIG performs a post-hoc analysis to identify the tools required for the target ELT workflow, as shown in Figure~\ref{fig:tool_flow}. kRAIG formulates a retrieval query to a RAG component to obtain tools relevant to the current task. The retrieved tool specifications (e.g., interfaces, parameters, and constraints) are then used as context for pipeline generation.
Transformation steps in the ELT sequence can be highly diverse, making it impractical to define a set of all required tools prior to generation. To address this, kRAIG is able to synthesize task-specific tools when existing capabilities are insufficient. However, automated tool synthesis introduces safety risks, as generated tools may unintentionally truncate, delete, or otherwise compromise data integrity. We discuss mitigation strategies for these failure modes in the safeguards section. To improve the reliability of tool selection and synthesis, we apply inference-time steering using example pipelines, which biases generation toward valid transformation implementations and reduces deviation from expected pipeline structure.

\section{Safeguards}

\begin{figure}[H]
    \centering
    \includegraphics[width=1\linewidth]{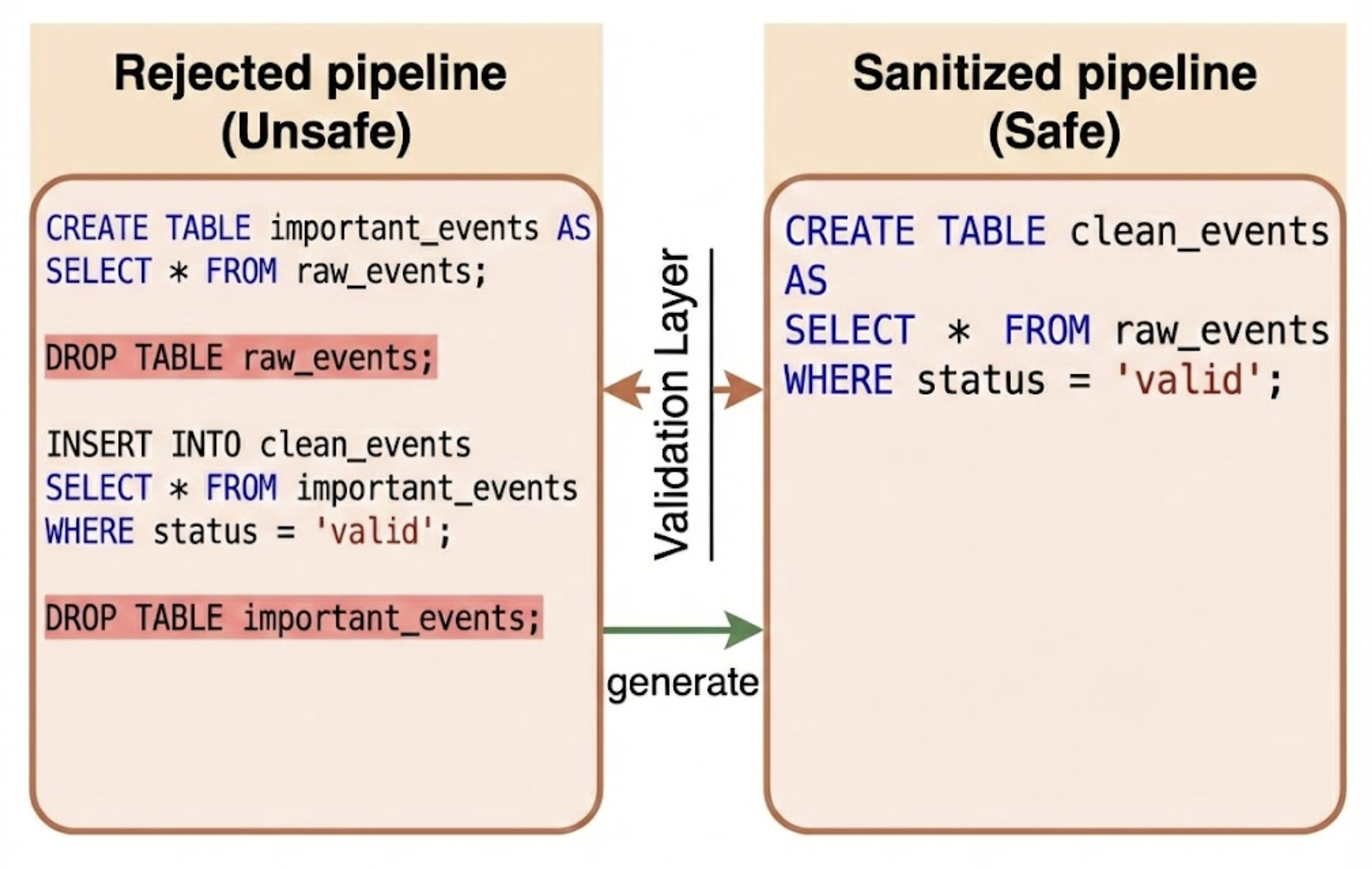}
    \caption{kRAIG LLM Validation}
    \label{fig:safeguard}
\end{figure}

Data security and safety are primary concerns in our design. Because kRAIG operates over potentially sensitive data lakes, we prevent the agent from issuing harmful operations. Prior work has shown that agentic systems may generate unsafe or destructive actions when operating without sufficient constraints.

To mitigate these risks, we implement two complementary safeguards:

\begin{itemize}
    \item \textbf{LLM validation of generated pipelines.}
    After pipeline synthesis, a validation stage inspects the resulting specification for destructive operations (e.g., DROP TABLE, bulk deletions, or other data-removal patterns), as illustrated in Figure~\ref{fig:safeguard}. Pipelines that violate safety constraints are rejected or automatically sanitized prior to execution.
    \item \textbf{Constrained tool usage.}
    For high-risk operations, kRAIG is restricted to a curated set of tools with enforced safety constraints, such as read-only access or scoped writes.
 The agent cannot arbitrarily construct or invoke low-level operations in these cases. Instead, it must compose workflows from these controlled tools, enforcing constrained and safe pipelines.
\end{itemize}
Together, these mechanisms reduce the likelihood that kRAIG can generate or execute pipelines that compromise data integrity.

\subsection{Kubeflow Pipeline Example}

\begin{figure}[H]
    \centering
    \includegraphics[width=0.9\linewidth]{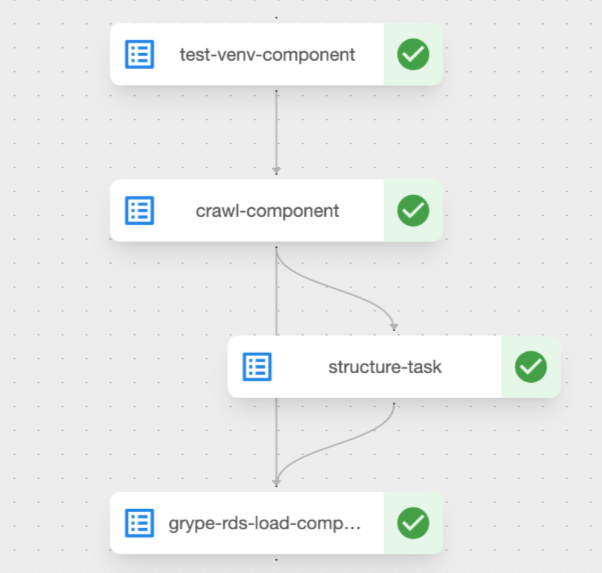}
    \caption{Example of a kRAIG-generated Kubeflow run}
    \label{fig:kraig_run}
\end{figure}

Figure~\ref{fig:kraig_run} shows an example execution of a pipeline generated by kRAIG within Kubeflow. The run illustrates how the synthesized pipeline is compiled and executed within the orchestration framework.

\section{Analysis and Data Visualization}
\begin{figure}[H]
    \centering
    \includegraphics[width=1\linewidth]{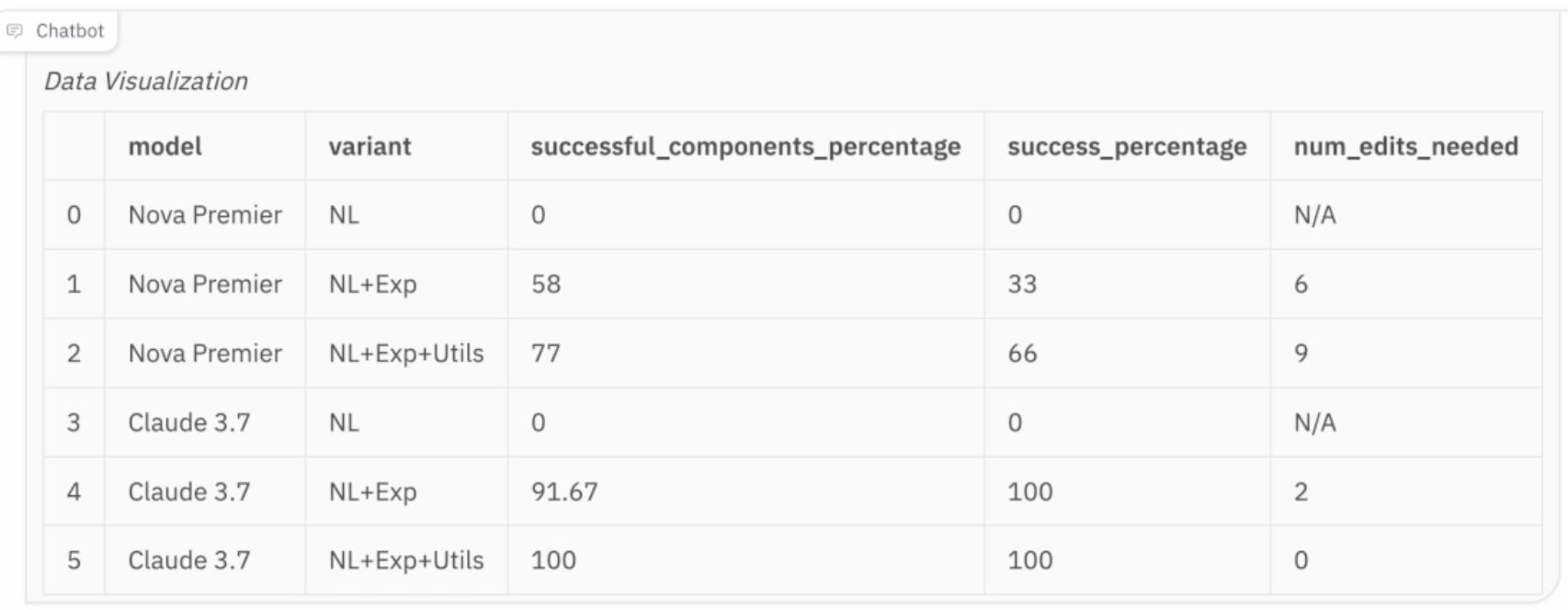}
    \caption{Example of kRAIG Generated Table}
    \label{fig:kraig_table_gen}
\end{figure}

\begin{figure}[H]
    \centering
    \includegraphics[width=1\linewidth]{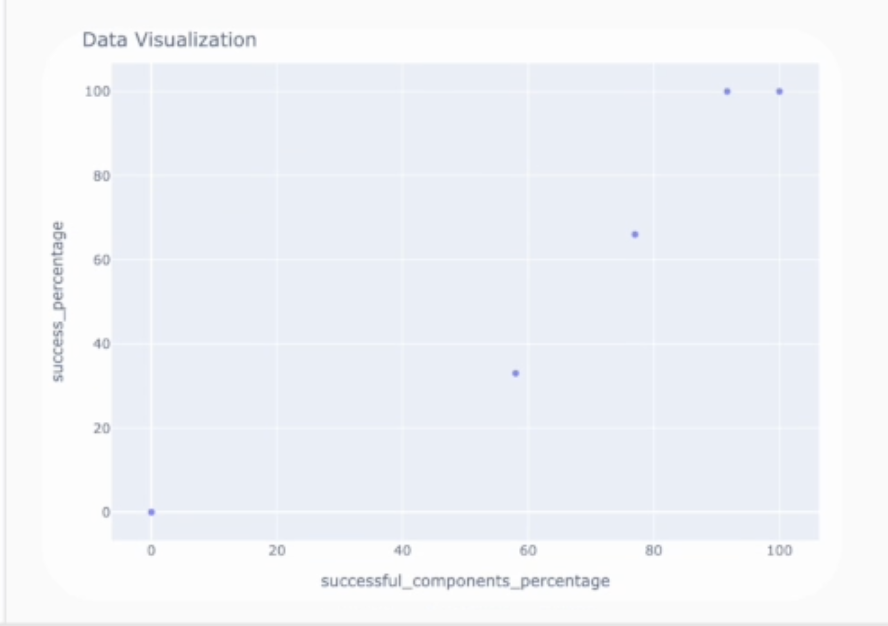}
    \caption{Example of kRAIG generated graph}
    \label{fig:kraig_graph_gen}
\end{figure}

We also extended kRAIG with capabilities for simple data visualization and aggregation tables, enabling effective auditing of the transformations performed within kRAIG’s pipeline. Examples of kRAIG-generated tables and visualizations on user-provided data are shown in Figures~\ref{fig:kraig_table_gen} and~\ref{fig:kraig_graph_gen}. kRAIG requires data extraction to be performed as a separate step prior to downstream augmentation and visualization.
For most tasks, kRAIG automatically generates a set of standard components:
\begin{itemize}[noitemsep, topsep=0pt]
    \item Data extraction
    \item Data transformation
    \item Data migration to the target source
    \item Basic data validation (e.g., comparing the number of rows inserted versus extracted)
\end{itemize}
These components enable users to audit intermediate results and verify correctness without manual checks. 

\subsection{Example of visualizations}

\section{Evaluation}

We evaluate kRAIG across three complementary dimensions: (i) model and prompting variant performance for pipeline generation, (ii) pipeline generation stability and variance, and (iii) end-to-end extract, loading and transform (ELT) performance on ELT Bench. Together, these evaluations assess kRAIG’s ability to generate executable pipelines, minimize human intervention, support complex transformations, and balance reproducibility.

\subsection{Model Variant Performance}

\begin{table*}
\centering
\begin{tabular}{llccc}
\hline
\textbf{Model} & \textbf{Variant} & \textbf{SC (\%)} & \textbf{SPC (\%)} & \textbf{Num Edits Needed} \\
\hline
Nova Premier & NL & 0 & 0 & N/A \\
Nova Premier & NL + Exp & 58 & 33 & 6 \\
Nova Premier & NL + Exp + Tools & 77 & 66 & 9 \\
\hline
Claude 3.7 & NL & 0 & 0 & N/A \\
Claude 3.7 & NL + Exp & 91.67 & 100 & 2 \\
Claude 3.7 & NL + Exp + Tools & 100 & 100 & 0 \\
\hline
\end{tabular}
\caption{\label{tab:model_variants}
Performance comparison across models and prompting variants. SC = Successful Components, SPC = Successful Pipelines Compiled.
}
\end{table*}

Table~\ref{tab:model_variants} evaluates two large language models, Nova Premier and Claude~3.7, under three prompting variants: NL (pure natural language), NL + Exp (natural language with grounding examples of valid Kubeflow pipelines), and NL + Exp + Tools (examples plus optional pre-set tools). Each configuration is evaluated over three independent runs.

We report three metrics:
\begin{itemize}
    \item \textbf{Successful Components (SC)}: average percentage of pipeline components that successfully compile across runs. The total number of components varies by task (typically 1--4).
    \item \textbf{Successful Pipeline Components (SPC)}: fraction of runs in which the entire pipeline compiles and executes end-to-end.
    \item \textbf{Num Edits Needed}: number of manual code edits required by a human to achieve a successful end-to-end pipeline.
\end{itemize}

\paragraph{Results.}
With pure natural language prompting, both models fail to generate executable pipelines, resulting in 0\% SC and 0\% SPC. In these cases, generated code does not conform to Kubeflow pipeline requirements and never reaches a state where manual edits are meaningful.

Adding grounding example pipelines (NL + Exp) substantially improves pipeline performance. Nova Premier achieves 58\% SC and 33\% SPC, while Claude~3.7 reaches 91.67\% SC and 100\% SPC with only two manual edits. This demonstrates that structural pipeline grounding is critical for mapping user intent to executable pipelines.

Providing optional tools (NL + Exp + Tools) further improves component-level and pipeline-level success. Claude~3.7 achieves 100\% SC and 100\% SPC with zero manual edits, indicating fully autonomous pipeline generation. Nova Premier also improves to 77\% SC and 66\% SPC, with the cost of more manual edits. This result highlights a capability–complexity tradeoff, where additional tooling increases expressiveness but also raises the rate of error for less capable models.

Overall, these results show that kRAIG’s design strongly benefits from structured prompting, and that model capability determines whether added tooling reduces or increases human intervention.

\subsection{Pipeline Generation Variance and Reproducibility}

\begin{table*}
\centering
\resizebox{\textwidth}{!}{%
\begin{tabular}{lccccccccc}
\hline
\textbf{Data Source} & \textbf{Avg Sim} & \textbf{Min Sim} & \textbf{Median Sim} & \textbf{Std Sim} & \textbf{Variance} & \textbf{Unique Versions} & \textbf{Duplication Gini} \\
\hline
HuggingFace Cauldron & 0.9802 & 0.9344 & 0.9875 & 0.0186 & 0.0198 & 14 & 0.2286 \\
HuggingFace Docmatix & 0.9978 & 0.9924 & 0.9968 & 0.0022 & 0.0022 & 6 & 0.5333 \\
HuggingFace Mathvision & 0.9973 & 0.9933 & 0.9973 & 0.0015 & 0.0027 & 15 & 0.2133 \\
Grype Github Repo & 0.9875 & 0.9237 & 0.9952 & 0.0197 & 0.0125 & 18 & 0.0889 \\
Numpy Github Repo & 0.9614 & 0.7949 & 0.9803 & 0.0474 & 0.0386 & 20 & 0 \\
YOLO Github Repo & 0.8990 & 0.7810 & 0.9007 & 0.0298 & 0.1010 & 20 & 0 \\
\hline
\end{tabular}%
}
\caption{\label{tab:data_sources}
Similarity and duplication statistics across data sources, computed over 20 runs using the same prompt, highlighting output consistency and redundancy.}
\end{table*}

We additionally evaluate kRAIG for generation stability under repeated prompting. Table~\ref{tab:data_sources} evaluates kRAIG’s pipeline generation variance across six data sources, including Hugging Face datasets and GitHub repositories. For each source, we generate 20 pipelines using identical prompts and compute pairwise similarity between generated pipelines using a normalized sequence-based string similarity metric over the serialized pipeline code, with aggregate similarity statistics used to quantify generation variance.

We report average, minimum, and median similarity, as well as variance, number of unique pipeline versions, and duplication Gini coefficient.

\paragraph{Results.}
Across most data sources, kRAIG exhibits high average similarity (0.96--0.99), indicating strong stability. Hugging Face datasets, which typically involve more standardized data access patterns, show particularly high similarity and low variance.

GitHub repositories requiring nontrivial transformations (e.g., YOLO and NumPy) exhibit higher variance, lower minimum similarity, and more unique pipeline versions. This behavior is expected and desired, as transformation-heavy pipelines allow multiple valid implementations, giving kRAIG flexibility in writing custom transformation logic. Importantly, even in these cases, pipelines remain executable and diverse rather than collapsing to a single template.

These results demonstrate that kRAIG balances stability and diversity, producing consistent pipelines for standardized tasks while allowing controlled variation when task complexity increases.

\subsection{ELT Bench Evaluation}

\begin{table*}
\centering
\begin{tabular}{llcc}
\hline
\textbf{Method} & \textbf{LLM} & \textbf{SRDEL (\%)} & \textbf{SRDT (\%)} \\
\hline
SWE-Agent & Claude-3.5-Sonnet & 12.5 & 0 \\
Spider-Agent & Claude-3.7-Sonnet (extended thinking) & 25 & 0 \\
Spider-Agent (Modified) & Claude-3.7-Sonnet (extended thinking) & \textbf{75} & 12.5 \\
\textbf{kRAIG (Ours)} & \textbf{Claude-3.7-Sonnet} & \textbf{75} & \textbf{25} \\
\hline
\end{tabular}
\caption{\label{tab:elt_bench} ELT Bench SRDEL and SRDT performance across agent frameworks and LLMs.}
\end{table*}

We further evaluate kRAIG on ELT Bench, an existing benchmark designed to measure extraction, loading, and transformation performance. We report two metrics:
\begin{itemize}
    \item \textbf{SRDEL}: success rate of extraction and loading.
    \item \textbf{SRDT}: success rate of transformations, measured across all runs.
\end{itemize}

Table~\ref{tab:elt_bench} compares kRAIG against SWE-Agent, Spider-Agent, and a custom modified variant of Spider-Agent adapted for ELT-style workflows.

\paragraph{Benchmark adaptation.}
ELT Bench was originally designed to evaluate general-purpose agents operating over a fixed set of data solutions. In contrast, kRAIG is designed as a specialized data engineering agent that leverages Python, Kubeflow, and complementary libraries to support diverse data sources, transformations, and target destinations.

To ensure a fair comparison, we randomly select eight ELT Bench tasks with evenly distributed difficulty, including simple to complex workflows. This same subset is used consistently to evaluate kRAIG, Spider-Agent, and SWE-Agent. Because ELT Bench tasks must be adapted to align with kRAIG’s deployment and execution constraints, we restrict evaluation to a limited subset. The selected eight tasks are sufficiently diverse to capture a broad range of workflow complexities while remaining representative of the benchmark as a whole. For baseline comparisons, we evaluate each agent using its strongest single model configuration, rather than averaging across multiple variants.

Importantly, while ELT Bench is designed to evaluate extraction, loading, and transformation, in practice it primarily exercises loading and transformation logic. kRAIG’s extraction capabilities extend beyond what the benchmark captures, including web scrapers, crawlers, Hugging Face dataset loaders, and custom data ingestion utilities. As a result, the benchmark should be interpreted primarily as an evaluation of load and transform performance.

\paragraph{Spider-Agent modification.}
In addition to the original Spider-Agent, we evaluate a modified variant adapted for ELT-style tasks. In practical data engineering settings, execution environments often vary and task descriptions frequently omit critical setup details (e.g., infrastructure configuration or schema assumptions). Under such ambiguity, purely iterative agents can fail due to repeated execution errors.

We modify Spider-Agent by updating its instruction examples to include a setup and structural details before pipeline generation. This change steers the agent toward more effective reasoning under under-specified instructions. We include this variant to reflect realistic prompt adaptation and to assess whether kRAIG’s results persist against a tuned baseline.

\paragraph{Results.}
Existing agents exhibit limited ELT performance. For each baseline, we report results using the best-performing model configuration identified during preliminary evaluation. SWE-Agent achieves an SRDEL of 12.5\%, with SRDT near zero across most model configurations. Even with extended reasoning, Spider-Agent reaches only 25\% SRDEL and 0\% SRDT, while the modified agent matches kRAIG's SRDEL with only a 12.5\% SRDT.

kRAIG achieves \textbf{75\% SRDEL} and \textbf{25\% SRDT}, substantially outperforming all baselines, particularly on transformation tasks where prior methods largely fail.

A key contributor to this improvement is kRAIG’s ability to clarify underspecified execution assumptions before pipeline generation. Several ELT Bench tasks involve MongoDB-backed workflows where critical configuration details (e.g., local vs. managed deployment or expected schema) are not specified in the prompt. 

In such cases, Spider-Agent and SWE-Agent must implicitly infer infrastructure details and frequently fail under repeated execution errors, leading to invalid pipelines. In contrast, kRAIG explicitly queries missing execution context before pipeline synthesis, preventing cascading failures and resulting in higher end-to-end success.

\subsection{Discussion}

Across all evaluations, kRAIG consistently demonstrates strong end-to-end performance:
\begin{enumerate}
    \item Structured prompting with grounding examples is essential for executable pipeline generation, while tool integration enables full autonomy for capable models.
    \item kRAIG produces stable and reproducible pipelines under repeated prompting, while allowing increased diversity for transformation-heavy tasks.
    \item On ELT Bench, kRAIG significantly improves on extraction, loading, and transformation performance compared to established agents.
\end{enumerate}

Beyond benchmarked tasks, kRAIG has been successfully applied internally to complex real-world workloads, including security data cleaning for CVE analysis and image preprocessing pipelines over Hugging Face datasets. These tasks involve diverse sources, nontrivial transformations, and heterogeneous destinations, highlighting kRAIG’s flexibility and practical utility.

These results validate kRAIG as a robust agent framework that reduces human intervention, supports complex transformations, and generalizes across a wide range of data engineering tasks.

\section*{Limitations}

kRAIG demonstrates the ability to generate data pipelines that consistently compile and execute. However, performance varies across task types. The system performs more reliably on extraction and load tasks than on complex transformation tasks, which often require more nuanced reasoning and domain-specific logic.

For extraction tasks, difficulty increases when workflows involve a large number of highly heterogeneous or geographically dispersed data sources, which can overwhelm kRAIG’s ability to accurately infer task structure and dependencies. In complex scenarios, kRAIG may also fail to identify missing task specifications and may proceed without sufficient clarification, leading to incomplete or suboptimal pipelines.

These limitations reflect broader challenges in current large language models’ ability to identify under-specified requirements and generate effective clarifying questions, consistent with observations reported in QuestBench \cite{li2025questbenchllmsaskright}.

\section*{Conclusion}
Our development and evaluation of kRAIG reveal several key distinctions relative to agents evaluated in ELT-Bench:

\begin{itemize}
\item \textbf{Reason–Question–Action Framework.} Introducing an explicit questioning phase prior to action selection reduces task ambiguity and prevents cascading failures caused by underspecified instructions. Analysis of Spider-Agent trajectories indicates that certain failures arise from schema inconsistencies between local and cloud environments, which additional reasoning iterations alone cannot resolve.
\item \textbf{Context Distillation.} Summarizing and distilling important information from conversational history and task specifications improves success rates on complex workflows, particularly when instruction length approaches model context limits.
\item \textbf{Limitations of Traditional ReAct Frameworks.} Pure reason-then-act approaches are more susceptible to ambiguity in natural language instructions, often resulting in inefficient reasoning paths and increased execution steps before reaching valid solutions.
\end{itemize}

In this iteration, the kRAIG agent is primarily designed to reliably execute internal tasks. Our development and evaluation efforts focus on leveraging the Kubeflow platform and implementing functionalities that support our researchers by enabling greater focus on data analysis, while automating repetitive operations and reducing manual effort within the machine learning workflow.

\section*{Future Work}
Our evaluation highlights several limitations that motivate future work. A primary direction is a comprehensive evaluation of kRAIG across the full ELT-Bench task suite to better characterize its strengths and failure modes. In parallel, we aim to expand kRAIG’s functional capabilities while maintaining strict adherence to security and governance constraints required for internal deployment.

\subsection{Expansion of Questioning Ability}
An important direction for future work is improving kRAIG’s ability to identify missing or ambiguous task specifications and to generate targeted clarifying questions. We plan to explore a range of training and post-training techniques to enhance the base model’s ability to infer intermediate steps required for diverse data engineering tasks and to detect when critical information is absent.

Once missing steps are identified, the agent must generate context-aware questions that elicit the necessary details from users. While kRAIG demonstrates reduced performance in this area, the inherent structure of extract, load, and transform workflows enable reasonable performance on standard ELT tasks through few-shot generalization, retrieval-augmented examples, and constrained tool usage.

\subsection{Enhancements on Data Visualization and Analytics}
We plan to enhance kRAIG’s data visualization and analytics capabilities, with improvements to complex transformation pipelines serving as a key enabling factor. Strengthening transformation support will allow kRAIG to more effectively generate, audit, and analyze intermediate and final datasets, enabling end-to-end workflows that extract, transform, and visually summarize data in response to user queries.

\subsection{Guardrail Improvement} We acknowledge that the LLM-based validation and static patterns used here are necessary but insufficient for high-stakes production environments, as they may be susceptible to adversarial 'jailbreaking' or sophisticated prompt injection. Next-generation security for kRAIG must move beyond reactive guardrails toward formal verification of generated code and sand-boxed symbolic execution. These methods would allow the system to mathematically guarantee that a generated pipeline cannot perform unauthorized data deletions or out-of-scope writes before a single line of code is executed.

\section*{Ethical Considerations}
kRAIG is an AI agent that translates natural-language requests into executable data engineering pipelines. Because such pipelines may access sensitive data sources and can perform write operations, incorrect generations or misuse could cause harms including inadvertent disclosure of private data, creation of non-compliant derived datasets, corruption or loss of data through unsafe transformations, and operational disruptions (e.g., triggering expensive jobs or propagating erroneous outputs downstream).

We explicitly design kRAIG with safety considerations for DataOps settings. First, we incorporate static validation of generated pipelines to detect and reject destructive operations (e.g., patterns resembling bulk deletions or irreversible data removal) prior to execution. Second, kRAIG is constrained to compose workflows from a curated set of vetted tools rather than issuing arbitrary low-level actions, reducing the likelihood of unsafe behaviors and supporting organizational policies such as least-privilege access. In practical deployments, kRAIG should additionally be operated with standard governance controls including authentication/authorization, audit logging, and clear separation between development and production environments, with human review for high-risk changes.

Our evaluation focuses on pipeline executability and benchmark performance; it does not constitute comprehensive security testing or red-teaming against adversarial prompts. Future work includes more rigorous evaluation of safety guarantees under adversarial or ambiguous instructions, and stronger policy enforcement for write operations and dataset handling across heterogeneous sources and destinations.

\section*{Acknowledgements}
We thank our DevOps team for their support, Shiv Deshmukh, Tim Carlson and team. We want to particularly highlight Shiv Deshmukh specialized assistance for implementing and maintaining the Kubeflow platform and providing the core infrastructure required to deploy and operate the kRAIG agent. We thank David Windsor for his valuable contributions to the conceptual development and constructive feedback during the refinement of this work. We also acknowledge Craig Connors for his leadership and providing resources necessary to pursue this research direction.
\section*{Use of AI Assistants}
We used AI assistants for editing assistance and code debugging during the preparation of this work. The authors take full responsibility for all content.

\nocite{yang2024sweagent}
\nocite{airbyte2025}
\nocite{sagemaker2025}
\nocite{andukuri2024stargateteachinglanguagemodels}
\nocite{anthropic_claude3}
\nocite{Intelligence2024}
\nocite{lei2024spider2}
\nocite{wang2023survey_lm_dialogue}
\nocite{yao2022react}
\nocite{schick2023toolformer}
\nocite{qu2024tool_learning_survey}
\nocite{bisk2020experiencegroundslanguage}
\bibliography{anthology}
\bibliographystyle{acl_natbib}
\end{document}